# Superconductivity under high pressure in LaFeAsO

Hironari OKADA*, Kazumi IGAWA, Hiroki TAKAHASHI, Yoichi KAMIHARA[1], Masahiro HIRANO[1,2], Hideo HOSONO[1,2], Kazuyuki MATSUBAYASHI[3], and Yoshiya UWATOKO[3]

*Department of Physics, College of Humanities and Sciences, Nihon University, Sakurajosui, Setagaya-ku, Tokyo 156-8550, Japan*

[1]*ERATO-SORST, JST, in Frontier Research Center, Tokyo Institute of Technology, 4259 Nagatsuta, Midori-ku, Yokohama 226-8503, Japan*

[2]*Frontier Research Center, Tokyo Institute of Technology, 4259 Nagatsuta, Midori-ku, Yokohama 226-8503, Japan*

[3]*Institute for Solid State Physics, The University of Tokyo, Kashiwanoha, Kashiwa, Chiba 277-8581, Japan*

Electrical resistivity measurements under high pressures up to 29 GPa were performed for oxypnictide compound LaFeAsO. We found a pressure-induced superconductivity in LaFeAsO. The maximum value of $T_c$ is 21 K at ~12 GPa. The pressure dependence of the $T_c$ is similar to those of $LaFeAsO_{1-x}F_x$ series reported previously.

KEYWORDS: superconductivity, high pressure, electrical resistivity, oxypnictide, LaFeAsO

---

*Corresponding author: Department of Physics, College of Humanities & Sciences, Nihon University
3-25-40 Sakurajosui, Setagaya-ku, Tokyo 156-8550, Japan
E-mail address: hironari@chs.nihon-u.ac.jp

Recent discovery of Fe-based oxypnictide superconductor $LaFeAsO_{1-x}F_x$ with superconducting transition temperature $T_c$ of ~26 K has attracted great interests in the field of condensed matter physics.[1] The crystal structure of $LaFeAsO_{1-x}F_x$ is a tetragonal ZrCuSiAs-type structure ($P4/nmm$), and is composed of a stack of insulating LaO and conducting FeAs layers. It is thought that conductive carriers are confined two dimensionally in the FeAs layers and the carrier concentration can be increased by the substitution of $O^{2-}$ by $F^{-}$. The parent compound LaFeAsO is metallic and undergoes a structural phase transition at ~160 K and a magnetic phase transition at ~140 K.[2,3] The crystal structure changes from the tetragonal to orthorhombic, then the compound exhibits a spin density wave type antiferromagnetic (AFM) ordering. The structural distortion and the magnetic ordering are suppressed by the F substitution, resulting in the emergence of the superconductivity.

Right after the discovery of the superconductivity in $LaFeAsO_{1-x}F_x$, it was reported that the $T_c$ of $LaFeAsO_{0.89}F_{0.11}$ shows large pressure effect and increases to a maximum of 43 K at 4 GPa, suggesting that the lattice contraction is effective in the enhancement of the $T_c$ of oxypnictide compounds.[4] Actually, $R$FeAsO ($R$ = rare earth) having smaller $R$ atom than La atom shows the $T_c$ over 50 K.[5,6] It is known that external pressure can control carrier concentration in $CuO_2$ plane of high-$T_c$ cupper oxides.[7] In order to understand the effect of external pressure on $T_c$ in $LaFeAsO_{1-x}F_x$ and clarify similarities and differences between the external pressure effect and the F substitution effect, we have performed electrical resistivity measurements under high pressures up to 29 GPa for the parent compound LaFeAsO. In this paper, we report a pressure-induced superconductivity in LaFeAsO with a maximum $T_c$ of 21 K.

A polycrystalline sample was prepared by a solid state reaction method.[1,2] The sample was confirmed to be mostly composed of a single phase of the ZrCuSiAs-type tetragonal structure with lattice constants of $a$ = 0.403268(1) nm and $c$ = 0.874111(4) nm at room temperature.[2] Electrical resistivity measurements under high pressures up to 1.5 and 12 GPa were performed by a standard dc four-probe technique using a piston-cylinder type pressure cell and a cubic anvil cell, respectively.[8] A liquid pressure-transmitting medium (Fluorinert FC70:FC77 = 1:1) was used to maintain hydrostatic conditions. A diamond anvil cell (DAC) made of CuBe alloy was used for electrical resistivity measurements under high pressures up to 29 GPa. The sample chamber comprising a rhenium gasket was filled with powdered NaCl as a pressure-transmitting medium.[4] AC magnetic susceptibility measurements with AC field of 1.8 mT and a frequency of 317 Hz were carried out using the cubic anvil cell.

Figure 1 shows the temperature dependence of the electrical resistivity of LaFeAsO using the piston-cylinder type pressure cell. The electrical resistivity of LaFeAsO shows a rapid decrease below ~150 K at 0 GPa.[1)] This anomaly at ~150 K becomes broader with applying pressure. The temperature dependence of $d\rho/dT$ shows a peak at 145 K at 0 GPa, and the peak of $d\rho/dT$ shifts to lower temperature side by applying pressure, as shown in the inset of Fig.1. From the results of powder X-ray diffraction measurement and nuclear magnetic resonance using the same batch sample as ours,[2,9)] it was reported that the structural phase transition from the tetragonal to the orthorhombic (*Cmma*) and the AFM ordering occur at ~165 and 142 K, respectively. The temperature $T_0$ at which $d\rho/dT$ shows the peak at 0 GPa is close to the AFM ordering temperature rather than the structural phase transition temperature. Therefore, the pressure dependence of the $T_0$ may be related to a suppression of the AFM ordering by applying pressure. Another remarkable anomaly of the electrical resistivity was observed under high pressure at lower temperatures. At 1.5 GPa, the electrical resistivity gradually decreases on cooling below ~10 K.

Figure 2 shows the temperature dependence of the electrical resistivity of LaFeAsO under high pressures up to 12 GPa, using the cubic anvil cell. The characteristic temperature $T_0$ decreases with applying pressure. On the other hand, the anomalous behavior at lower temperatures becomes more distinct. Although the electrical resistivity at the lowest temperature shows a finite value in the lower pressure regions, the decrease in the electrical resistivity becomes steeper with increasing pressure, and finally, we observed a zero resistivity at 12 GPa.

Figures 3(a) and 3(b) show the magnetic field effect at 7.5 GPa and the current effect at 12 GPa on the electrical resisitivity below 40 K, respectively. The temperature at which the electrical resistivity begins to decrease, indicated by arrows in Fig. 3(a), shifts to the lower temperature and the electrical resistivity at 0 T below the temperature indicated by the arrows increases by applying magnetic field. The zero resistivity observed at 12 GPa disappears by increasing current density, as shown in Fig. 3(b). Thus, these results indicate that the sudden decrease of the electrical resistivity is reduced by applying magnetic field and by increasing current density. Moreover, a diamagnetic signal of AC magnetic susceptibility is detected below ~20 K at 9 and 12 GPa, as shown in Fig 3(c). In this way, the magnetic field and the current effects on the electrical resistivity and the diamagnetic signal strongly suggest that the anomalous behavior in the electrical resistivity below 20 K is attributed to a pressure-induced superconductivity. The broadening of the superconducting transition below ~20 K is probably due to a sample inhomogeneity rather than a distribution of pressure, because a hydrostatic condition was kept unchanged during the measurements. The pressure-induced superconductivity have also been reported in Fe-As compounds having a $ThCr_2Si_2$-type tetragonal structure.[10-16)]

The anomaly due to the pressure-induced superconductivity is also detected by the electrical resistivity measurements under high pressures up to 29 GPa using the DAC, as shown in Fig. 4. In this measurement, the zero resistivity was not observed even in the highest pressure. The failure to observe the zero resistivity is probably caused by a nonhydrostatic compressive stress resulting from the use of a solid pressure-transmitting medium. We confirmed current dependence of the electrical resistivity below the transition temperature. The transition temperature increases by applying pressure to a maximum of 21 K, and decreases monotonically above 16 GPa.

Figure 5(a) shows the pressure dependence of the characteristic temperature $T_0$ and the superconducting transition temperature $T_c$. The $T_0$ monotonically decreases with increasing pressure. On the other hand, the $T_c$ increases to the maximum of 21 K at ~12 GPa, then deceases with further applying pressure. This *P*-*T* phase diagram shown in Fig. 5(a) is similar to *x*-*T* phase diagram for the $F^-$ substituted system, $LaFeAsO_{1-x}F_x$.[1)] In addition, the pressure dependence of $T_c$ in LaFeAsO is similar to those observed in the F-doped compounds, as shown in Fig. 5(b).[4,15)]

The substitution effect of $O^{2-}$ by $F^-$ in LaFeAsO is expected to lead shrinkage of lattice constants and change of carrier density, resulting in suppression of structural and magnetic phase transitions and enhancement of $T_c$. By the $F^-$ substitution of 5 at%, the unit cell volume changes from 0.14233 $nm^3$ to 0.14186 $nm^3$ and the superconductivity is induced at 24 K.[1)] Considering the pressure dependence of unit cell volume of F-doped compounds,[15)] this volume contraction corresponds to the application of external pressure of ~0.3 GPa. Therefore, from the viewpoint of volume contraction, the F substitution is more effective in the suppression of structural and magnetic phase transitions and in the emergence of superconductivity than the external pressure effect. Moreover, LaFeAsO shows the lowest maximum $T_c$ in $LaFeAsO_{1-x}F_x$ series under high pressure. The maximum $T_c$ of $LaFeAsO_{0.95}F_{0.05}$ is rather lower value of 29 K under high pressure, compared with the maximum $T_c$ of 43 K under high pressure in $LaFeAsO_{0.89}F_{0.11}$ and $LaFeAsO_{0.86}F_{0.14}$. This may be due to a magnetic instability existing close to superconducting phase. On the other hand, the maximum $T_c$ of 43 K in “over-doped” $LaFeAsO_{0.86}F_{0.14}$ is comparable to that in “optimum-doped” $LaFeAsO_{0.89}F_{0.11}$ although the $T_c$ of $LaFeAsO_{0.86}F_{0.14}$ at 0 GPa is smaller than that of $LaFeAsO_{0.89}F_{0.11}$ by 9 K. This result is in strong contrast to the case of high-$T_c$ cupper oxides. In high-$T_c$ cupper oxides, external pressure effect has an aspect of carrier doping effect, because carrier concentration in the $CuO_2$ layers increases by pressure through changing charge distribution due to anisotropic compression. In this case, the $T_c$ of "over-doped" $LaFeAsO_{0.86}F_{0.14}$ decreases with increasing external pressure.[18] Therefore, it is considered that the effect of external pressure on the electronic state of Fe-based superconductors is different with the case of high-$T_c$ cupper oxides.

In summary, we performed electrical resistivity measurements under high pressures up to 29 GPa for oxypnictide compound LaFeAsO. The anomaly due to the structural phase transition and/or the AFM ordering becomes broader with applying pressure. On the other hand, the pressure-induced superconductivity was discovered. The superconducting transition temperature increases to the maximum of 21 K at ~12 GPa. The further high-pressure studies for LaFeAsO are now in progress, in order to clarify the pressure effects on the structural phase transition, the AFM ordering, and the pressure-induced superconductivity in LaFeAsO.

**Acknowledgents**

The electrical resistivity measurements using cubic anvil cell were carried out by the joint research in the Institute for Solid State Physics, the University of Tokyo.

**Figure caption**

Fig. 1. (Color Online) Temperature dependence of the electrical resistivity under various pressures up to 1.5 GPa, using the piston-cylinder type pressure cell. The inset shows the temperature dependence of $d\rho/dT$ at 0 and 1.0 GPa.

Fig. 2. (Color Online) Temperature dependence of the electrical resistivity under various pressures up to 12 GPa, using the cubic anvil cell.

Fig. 3. (Color Online) (a) Temperature dependence of the electrical resistivity under 7.5 GPa at magnetic field $B$ = 0 and 3 T. (b) Temperature dependence of the electrical resistivity under 12 GPa at current density $j$ = 1 and 10 kA/m$^2$. (c) Temperature dependence of the AC magnetic susceptibility at 12 GPa. These data were obtained using the cubic anvil cell.

Fig. 4. (Color Online) Temperature dependence of the electrical resistivity below 40 K under various pressures up to 29 GPa, using the DAC.

Fig. 5 (Color Online) (a) The $P$-$T$ phase diagram of LaFeAsO obtained by the electrical resistivity measurements. Triangles and circles shows the characteristic temperature $T_0$ determined at the peak of $d\rho/dT$ and superconducting transition temperature $T_c$ determined at the temperature at which the electrical resistivity begins to decrease below ~20 K. (b) Pressure dependence of the superconducting transition temperature in $LaFeAsO_{1-x}F_x$ series. The solid curves are guides to the eye. The data of the F-doped compounds were reported in Ref. 4 and 15.

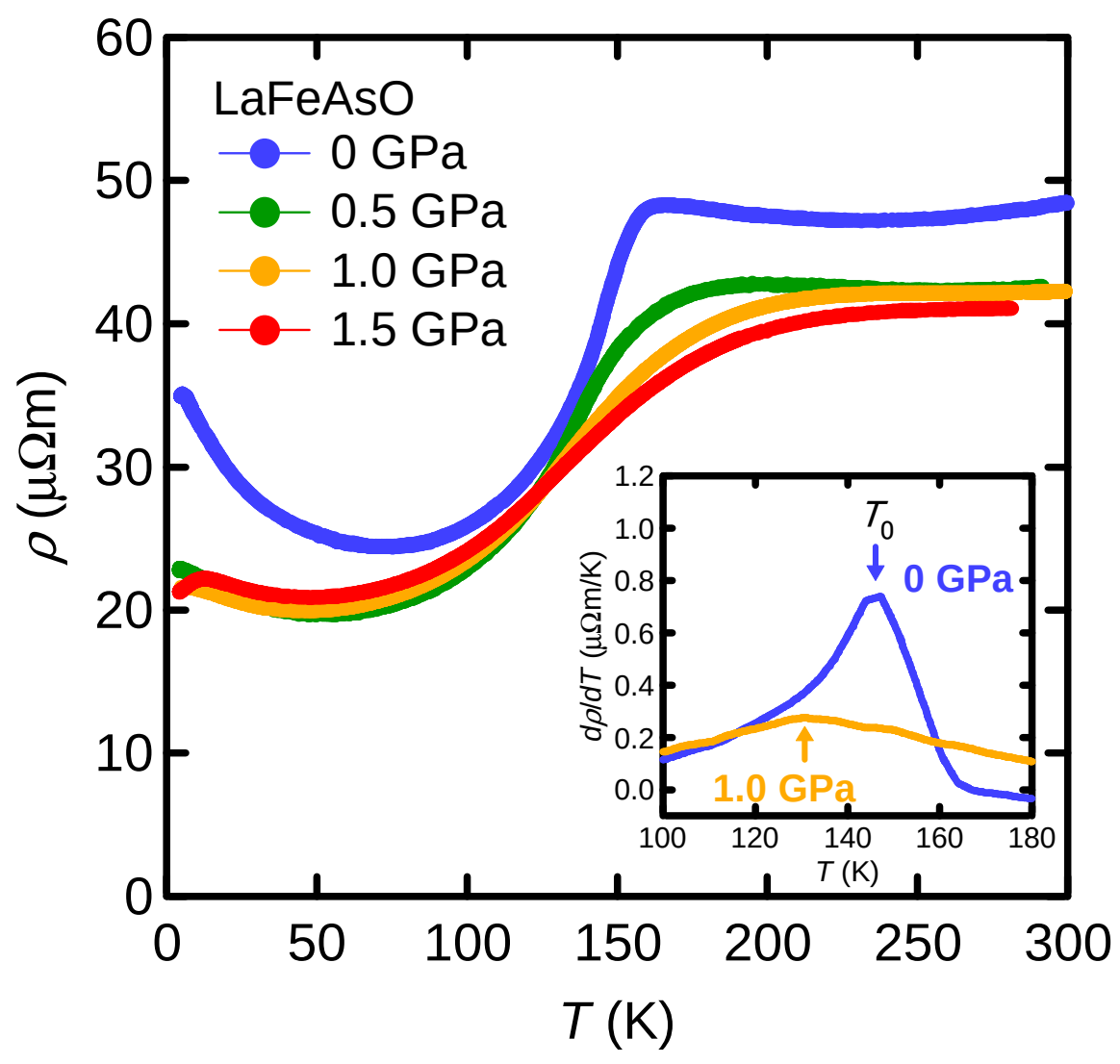

Fig. 1

H. Okada *et al*.

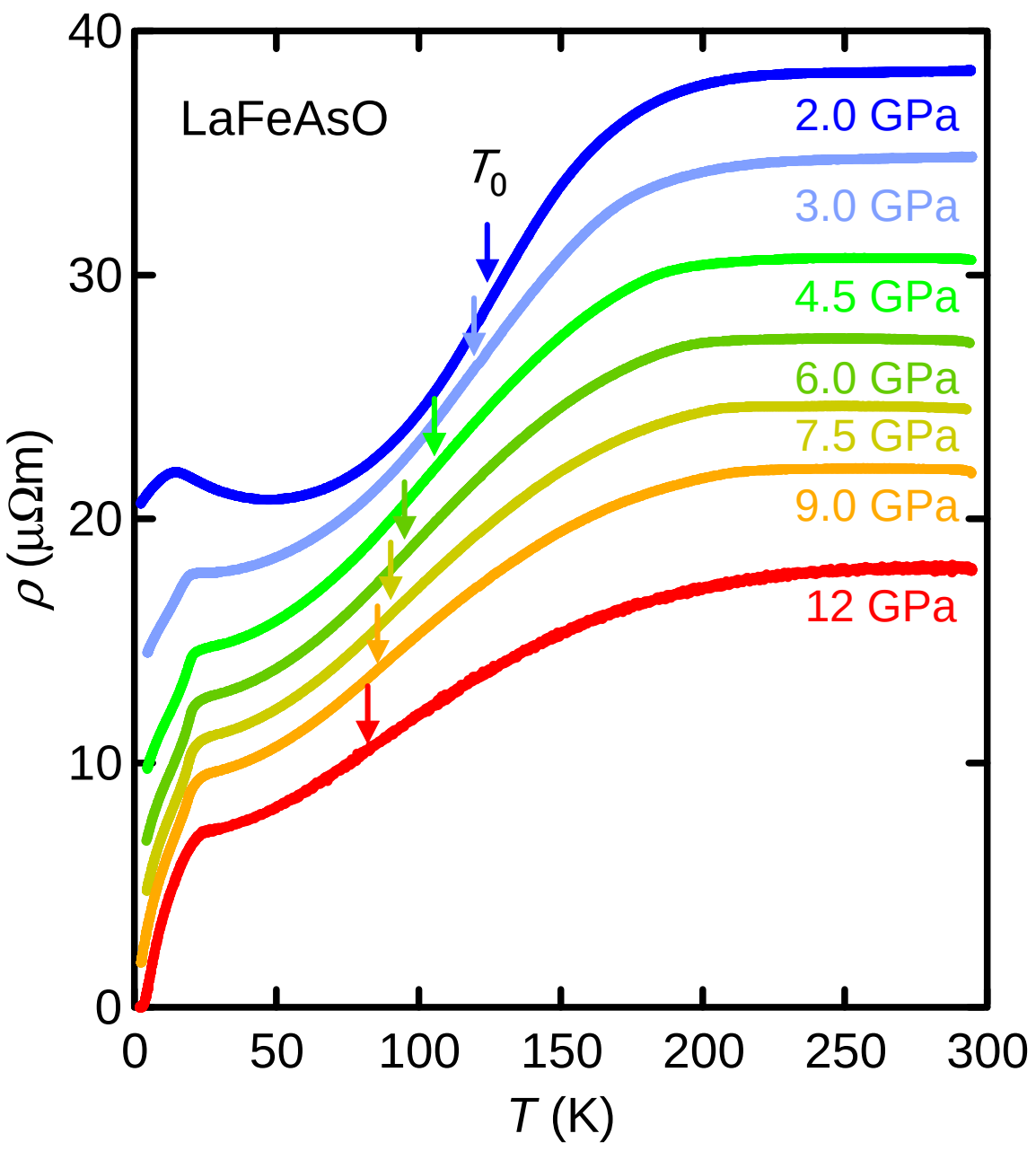

Fig. 2

H. Okada *et al*.

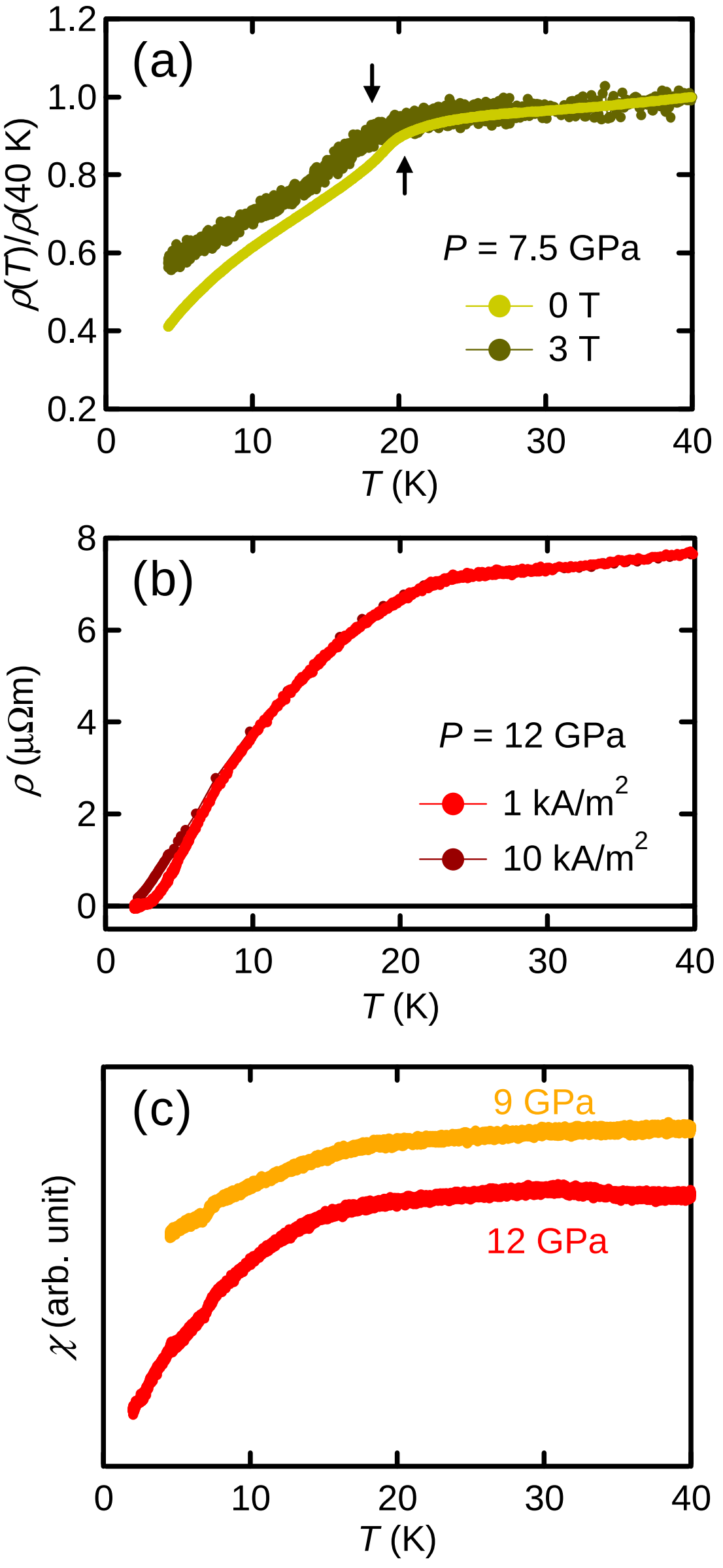

Fig. 3

H. Okada *et al.*

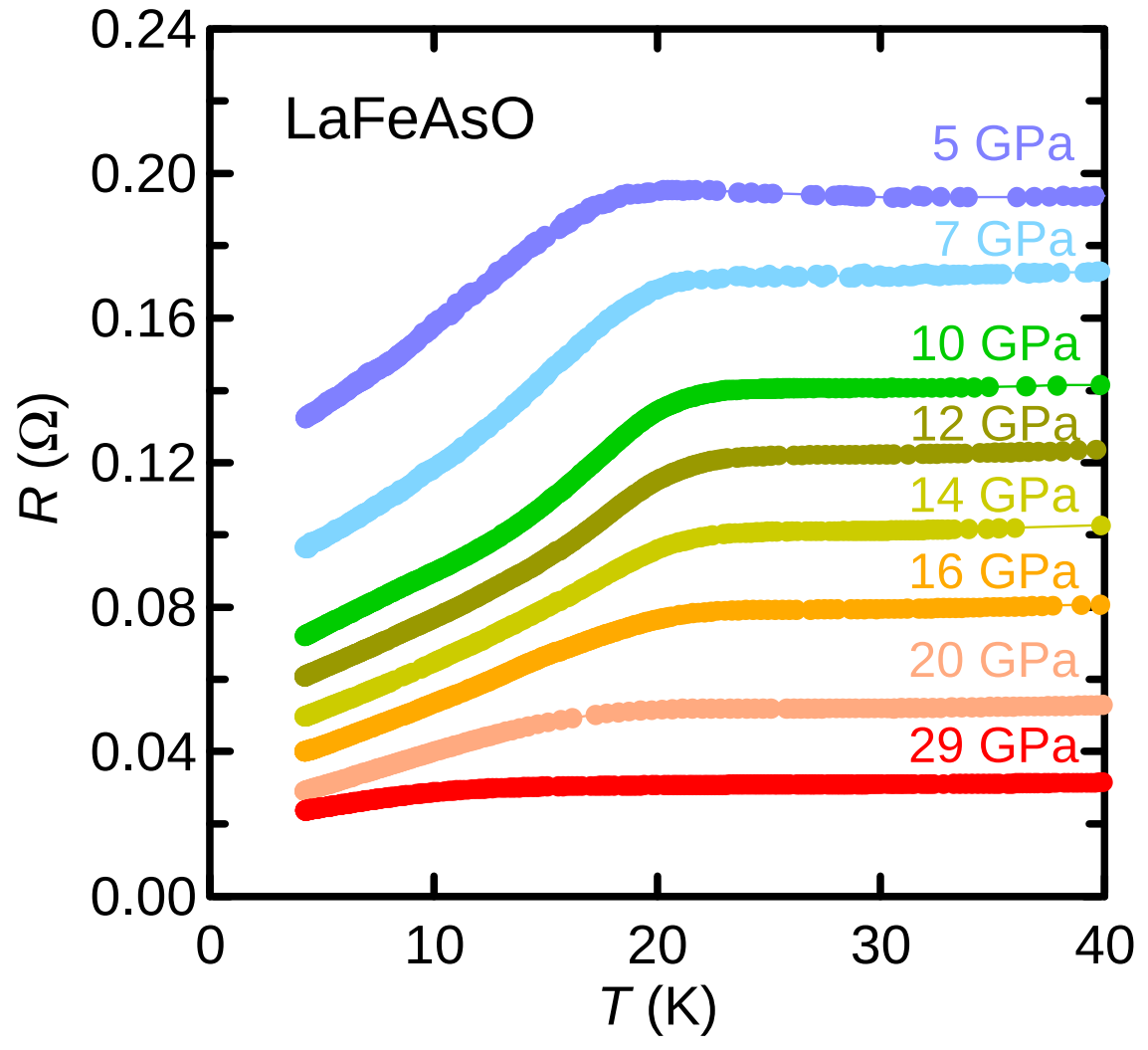

Fig. 4

H. Okada *et al*.

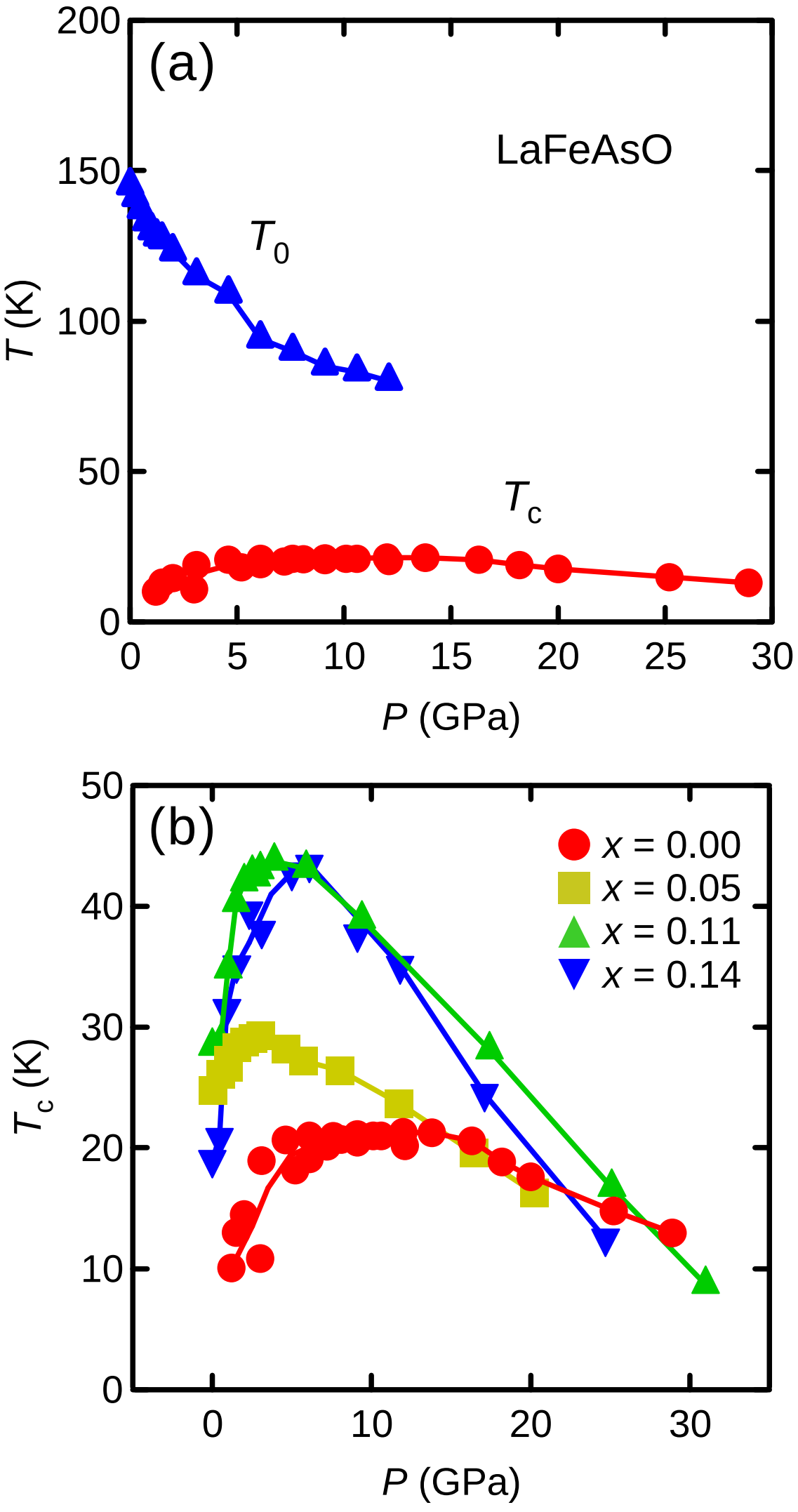

Fig. 5

H. Okada *et al*.